\documentclass[prb,twocolumn,amsmath,amssym,superscriptaddress,longbibliography]{revtex4-2}
\usepackage[T1]{fontenc}
\usepackage[latin9]{inputenc}
\usepackage{amsmath}
\usepackage{amsthm}
\usepackage{amssymb}
\usepackage{graphicx}
\usepackage{color}


\usepackage{babel}
\begin{document}

\title{Formation of spin and charge ordering in the extended Hubbard model during a finite-time quantum quench }

\author{Isaac M. Carvalho}%
\affiliation{Departamento de F\'isica, Universidade Federal de Minas Gerais,
  C.P.~702, 30123-970, Belo Horizonte, MG, Brazil}
\author{Helena Bragan\c{c}a}
\affiliation{Instituto de F\'{i}sica and International Center for Physics, Universidade de Bras\'{i}lia, Bras\'{i}lia 70919-970, DF, Brazil}%
\author{Walber H. Brito}
\affiliation{Departamento de F\'isica, Universidade Federal de Minas Gerais,
 C.P.~702, 30123-970, Belo Horizonte, MG, Brazil}
\author{Maria C. O. Aguiar}
\affiliation{Departamento de F\'isica, Universidade Federal de Minas Gerais,
  C.P.~702, 30123-970, Belo Horizonte, MG, Brazil}

\date{\today}

\begin{abstract}

We investigate the formation of charge and spin ordering by starting from a non-interacting state and studying how it evolves in time under a Hamiltonian with finite electronic interactions. We consider the one-dimensional, half-filled extended Hubbard model, which we solve within time-dependent density matrix renormalization group. By employing linear finite-time quenches in the onsite and nearest-neighbor interactions, we find the existence of impulse, intermediate, and adiabatic regimes of time evolution. 
For the quenches we analyze, we observe that the adiabatic regime is reached with distinct ramping time scales depending on whether the charge density wave (CDW) or the spin density wave (SDW) is formed. 
The former needs to be slower than the latter to prevent entangled excited states from being accessed during the quench. 
More interestingly, in the intermediate regime, we observe an enhancement of the entanglement entropy with respect to its initial value, which precedes the formation of the CDW ordering; a similar enhancement is not seen in the quench towards SDW. 
Our findings also show that the breaking of the system integrability, by turning on the nearest-neighbor interactions, does not give rise to significant changes in the non-equilibrium behavior within the adiabatic approximation.

\end{abstract}

\maketitle

\section{Introduction}

The dynamics of closed many-body systems following a quantum quench has become an active topic of research. 
In solid-state experimental setups, realizing a platform that is, at the same time, sufficiently well isolated from the environment and accessible for the experimental probe is a hampering factor to study the coherent evolution after a perturbation of the system. 
Ultracold atoms trapped in optical lattices~\cite{optl1,optl5,optl4,optl2,optl3,optl6}, on the other hand, provide an unprecedented opportunity to explore non-equilibrium phenomena due to the large set of available methods to
isolate, manipulate and measure these systems~\cite{langen2015ultracold}. 
In this experimental framework, the system parameters can be tuned either abruptly (sudden quench)~\cite{kasztelan2011landau} or by a finite-time protocol~\cite{chen2011many,chen2011quantum}. Furthermore, ultracold atoms give access to novel observables and
extreme parameter regimes, that go beyond the ones accessible in solid-state systems.
Alkali atoms (such as potassium $^{40}$K and lithium $^6$Li) in optical lattices, for instance, can be used as quantum simulators of the Fermi-Hubbard model with highly controllable local interaction~\cite{tarruell2018quantum}. Recently, nearest-neighbor interactions have also been realized using Rydberg dressing of $^6$Li~\cite{guardado2021quench}. 
From the theoretical side, the investigation of quantum quenches can be done via Conformal Field Theories~\cite{Calabrese_2004,calabrese2005evolution,calabrese2006time, calabrese2007entanglement} and tensor-network algorithms~\cite{dmrg1,dmrg3,dmrg4,dmrg2}, however,  the investigations are hampered by the entanglement growth. The analysis of the out-of-equilibrium dynamics of strongly-correlated fermionic systems is still a challenging problem.

In this context, the non-equilibrium dynamics after a quench through a quantum critical point is especially interesting. It has been shown that systems whose low-energy spectrum is described by a conformal field theory feature a large degree of universality even for late times after crossing the critical point \cite{cardy2016quantum,calabrese2006time}. 
Furthermore, for some systems that cross the criticality, the crossover between adiabatic and non-adiabatic stages of time evolution has been successfully described by a quantum extension of the  Kibble-Zurek mechanism, which is based on equilibrium critical exponents~\cite{Dziarmaga,delCampo_2014}. 
However, the complete theoretical description of the dynamics of strongly correlated systems going through a quantum phase transition (QPT) constitutes an open problem in many situations.

The extended Hubbard chain is a prototypical model for QPTs in strongly correlated systems. Its equilibrium version, for repulsive interactions and at half-filling, features a spin density wave (SDW), a charge density wave (CDW), as well as a bond-order wave (BOW) insulating phases~\cite{EjimaEHM2007}. 
As the system is metallic in the absence of interactions, it goes through metal-insulator QPTs when the interactions are turned on. The study of CDW and SDW phases attracts increasing attention due to experimental evidences of their interplay with the superconducting phases of cuprates~\cite{Miao2021} and iron-pnictides~\cite{Yi2014}.
Moreover, recent ultrafast optical experiments have addressed the nonequilibrium CDW phases presented in different families of correlated transition-metal dichalcogenides. For instance, the physics related to the nonthermal melting of CDW ordering in TiSe$_2$~\cite{Mazin_TiSe2} and VTe$_2$~\cite{tetsu_vte2}  has triggered great interest and is still under debate, since a purely electron-phonon mechanism can not account for the CDW melt-state.
The extended Hubbard model with an attractive nearest-neighbor interaction is also of great interest. Recently, it was used to describe the photoemission spectrum of the one-dimensional cuprate Ba$_{2-x}$Sr$_{x}$CuO$_{3+\delta}$~\cite{ZXShen_Science2021}. According to Wang and co-workers~\cite{Wang_PRL2021}, an attractive interaction appears in this case due to a long-range electron-phonon coupling.

In the non-equilibrium scenario, it was observed that a transition between correlated
and uncorrelated states can be driven by a quench in an external field added to the
Hubbard model~\cite{zhang2016quench}. Further, the effects of many-body interactions on the
statistics of energy fluctuations were investigated in the inhomogeneous version of the model,
in which the system was submitted to an out-of-equilibrium transient current along the
chain~\cite{zawadzki2020work}. 
In the case of the extended Hubbard model, the non-equilibrium states generated by radiation pulses
were recently described by a generalized Gibbs ensemble~\cite{murakami2022exploring}; the
non-equilibrium phase diagram of the system includes a $\eta$-paring superconducting phase.

In this work, we investigate the formation of SDW and CDW ordering in the extended Hubbard model submitted to interacting quenches. The initial state is chosen to be a delocalized one, the ground state (GS) of the non-interacting Hamiltonian, and we let it evolve under an interacting Hamiltonian. The onsite and/or nearest-neighbor repulsive interactions increase linearly in time, such that the final Hamiltonian is within the correlated SDW or CDW phase. At equilibrium, in the thermodynamic limit, the system is metallic at the non-interacting point and is an insulator for any finite interaction (within the CDW phase, charge and spin excitations are gapped, while in the SDW phase spin excitations are gapless~\cite{EjimaEHM2007}). Therefore, we start the evolution from a model critical point and investigate the formation of ordered phases during the quench.

We observe different regimes by varying the duration of the quench - we go from sudden quenches to adiabatic ones. We start by turning on either the onsite or the nearest-neighbor interaction, aiming to reach the SDW and the CDW ordering, respectively. For the latter, the intermediate regime, that precedes the adiabatic one, is characterized by an increase of the entanglement entropy with respect to the initial state, indicating that the evolved state includes excited disordered ones. As a consequence, we observe that the adiabatic regimes are reached by distinct values of the ramping time scales, longer for the CDW case than for the SDW one. Interestingly, the inclusion of small onsite (nearest-neighbor) interactions - meaning that both interactions are now turned on - hardly changes the dynamics during the quench towards the CDW (SDW) phase.

The organization of the paper is the following: in Sec.~\ref{sec_model} we present the model and the quench protocol; details of the numerical calculations are also mentioned. Our results are presented in Sec.~\ref{sec_results}. In Sec.~\ref{sec_results1} we discuss the results obtained by turning on only one of the interactions, either the onsite $U$ or the nearest-neighbor $V$ interaction. We observe the state time evolution during the quench in different regimes: adiabatic, intermediate, and impulse ones. 
Later, in Sec.~\ref{sec_results3}, we discuss the effects of turning on both $U$ and $V$ simultaneously. Our conclusions are summarized in Sec.~\ref{sec_conclusion}. The dependence of our main results on the chain size are presented in Appendix A.

\section{Model and quench protocol} \label{sec_model}

We investigate the time-dependent extended Hubbard model (EHM), given by the Hamiltonian

\begin{eqnarray}
\hat{H}\left(t\right)&=&-J\sum_{j,\sigma}\left(a_{j,\sigma}^{\dagger}a_{j+1,\sigma}+H.c.\right)\nonumber \\
&+&U\left(t\right)\sum_{j}\hat{n}_{j\uparrow}\hat{n}_{j\downarrow}+V\left(t\right)\sum_{j}\hat{n}_{j}\hat{n}_{j+1}
\label{eq_EHM}
\end{eqnarray}
which considers nearest-neighbor hopping of amplitude $J$ and time-dependent
onsite and nearest-neighbor interactions, given by $U$$\left(t\right)$ and
$V\left(t\right)$, respectively. In the equation above, $a_{j,\sigma}^{\left(\dagger\right)}$
annihilates (creates) a fermion with spin $\sigma=\uparrow,\downarrow$
on lattice site $j$, $\hat{n}_{j,\sigma}=a_{j,\sigma}^{\dagger}a_{j,\sigma}$,
and $\hat{n}_{j}=\hat{n}_{j,\uparrow}+\hat{n}_{j,\downarrow}$.


\begin{figure*}
\begin{centering}
  \includegraphics[scale=0.315]{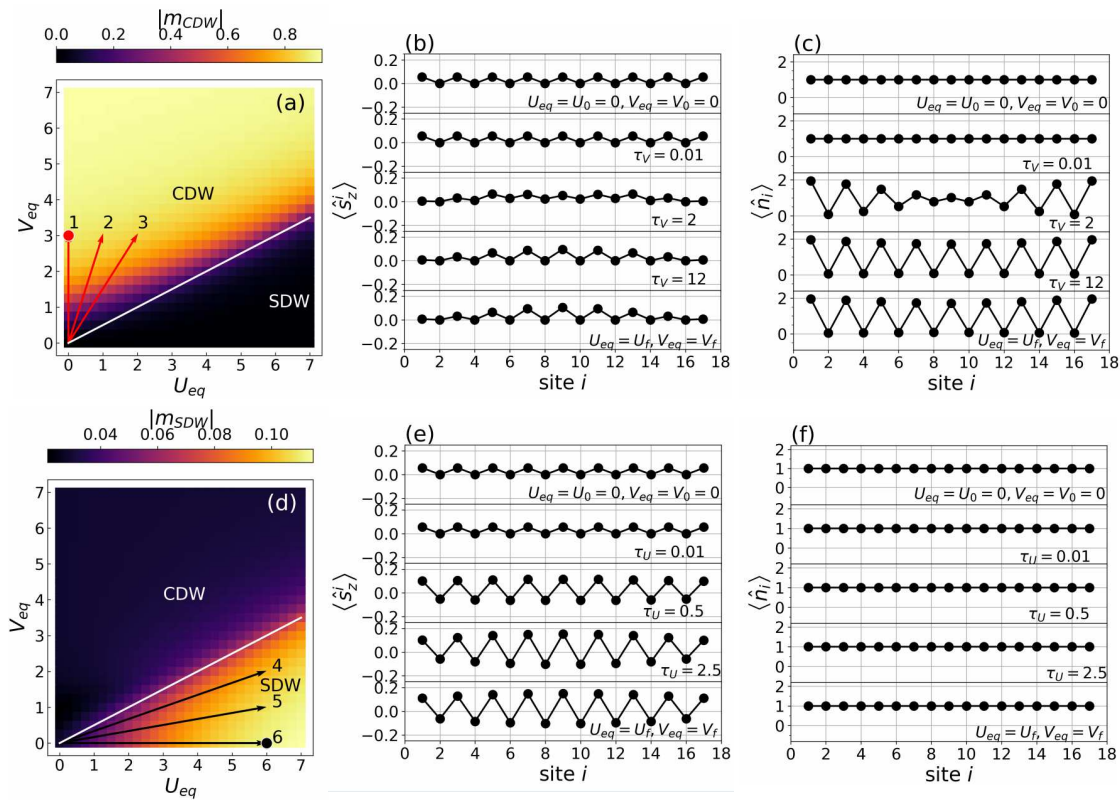}
\par\end{centering}
\caption{Equilibrium values for the (a) CDW and (d) SDW order parameters, as indicated by the color-codes, as a function of $U_{eq}$ and $V_{eq}$. 
The white straight lines correspond to $U=2V$.
The arrows indicate different quench directions to be explored along the paper. The red (black) dot identifies the final states corresponding to quench \#1 (\#6). The local magnetization and charge density as a function of the lattice site $i$ corresponding to the final state of \#1 quench are shown in panels (b) and (c), respectively, whereas the behavior of the final state of \#6 quench is depicted in panels (e) and (f). Besides showing results concerning the final state for quenches with different time scales ($\tau_V$ and $\tau_U$), we also present the profiles corresponding to the equilibrium GS with $U_{eq}=U_0=0$ and $V_{eq}=V_0=0$ and with $U_{eq}=U_f$ and $V_{eq}=V_f$. 
\label{EHM_phase_diagram}}
\end{figure*}

In this paper we study the time evolution of the system described by the
Hamiltonian of Eq.~(\ref{eq_EHM}) during a finite-time
quench in the interaction strengths. We first address the case in which only
one of the interactions, either onsite or nearest-neighbor one, is turned
on (Sec.~\ref{sec_results1}); later, we turn on both of them
simultaneously (Sec.~\ref{sec_results3}). We are
particularly interested in the formation of spin or charge ordered states and
we thus start from a delocalized state, that is, the GS of the Hamiltonian with no interactions
($U_{0}=V_{0}=0$). Then over a finite time
interval we evolve the parameters $U(t)$ or/and $V(t)$ from their initial values,
$U_{0}$ and $V_{0}$, up to their final values, $U_{f}$ and $V_{f}$. To be precise, we change the interactions linearly in time as
follows:

\begin{equation}
U\left(t\right)=U_{0}+sgn(U_{f}-U_{0})\frac{t}{\tau_{U}}
\end{equation}
or/and 
\begin{equation}
V\left(t\right)=V_{0}+sgn(V_{f}-V_{0})\frac{t}{\tau_{V}},
\end{equation}
where $t\in\left[0,t_{f}\right]$ and $\tau_{U}$ and $\text{\ensuremath{\tau_{V}}}$
are the ramping time scales. We set $t_{f}$ such that at the end of quench the values of the interactions are $U(t_{f})=U_{f}$ or/and $V(t_{f})=V_{f}$. If only one of the interactions is turned on,
for a final Hamiltonian within the SDW phase, we have $t_{f}=|U_{F}-U_{0}|\tau_{U}$ and $V_f=V_0=0$; for a final Hamiltonian in the CDW phase, we have $t_{f}=|V_{F}-V_{0}|\tau_{V}$ and $U_f=U_0=0$. Finally, when both interactions are turned on simultaneously, $\tau_U$ and $\tau_V$ are connected through $\tau_{V}=\tau_{U}|U_{F}-U_{0}|/|V_{F}-V_{0}|$.


Throughout this paper energies are given in units of $J$, and, accordingly, time is measured in units of $1/J$. The system is fixed at half-filling, the total magnetization in the $z-$direction is conserved during the quench, and we use open boundary conditions. Our results were obtained mainly for chains of size $L=17$, but we have confirmed our main findings by increasing the system length, as discussed in Appendix~\ref{sec_size}. 
The non-equilibrium numerical simulations of our global quench rapidly become computationally cumbersome with the number of sites.
In finite systems as those we simulate, the gap remains non-zero for all the Hamiltonian parameters. This quantity is key in the adiabatic theorem~\cite{messiah1999quantum} - it states that, in driven transitions to states with non-zero gaps, it is always possible to reach the adiabatic limit if the quench process is slow enough. Based on this theorem and on the finite $L$ of our systems, we expect to be able to observe an adiabatic behavior, as we indeed do (see the discussion of our results).

To obtain the system state at time $t$, we have performed time-dependent density matrix renormalization group (DMRG) calculations~\cite{dmrg2, dmrg3} with a first order Suzuki-Trotter decomposition, meaning that the error in the time evolution is of $\mathcal{O}\left(d\tau^{2}\right)$, where $d\tau$ is the time step. We consider $d\tau$ between $10^{-3}$ and $10^{-2}$, depending on the ramping time scale. Our implementation was built using the ITensor library~\cite{itensor}. In addition, for a better comparison with the evolved state during the quench, we have performed DMRG calculations to obtain the instantaneous equilibrium GS corresponding to constant $U_{eq} = U(t)$ and $V_{eq} = V(t)$ at each instant of time $t$. Our calculations were performed until the GS energy convergence was of the order of $10^{-8}$.

In the following, we present our numerical results for the system dynamics
during the quenches towards different regions of the phase diagram.
As we will show, we find significant differences in the relationship between the ramping time scale and
the adiabatic behavior depending on which ordering (CDW or SDW one) the final Hamiltonian corresponds to.

\section{Numerical results}\label{sec_results}

We start by obtaining the equilibrium phase diagram of the EHM within our implementation. To characterize both CDW and SDW phases, respectively, we use the order parameters defined as
\begin{equation}
  m_{CDW}=\frac{1}{L}\sum_{j}\left(-1\right)^{j}\left(\left\langle \hat{n}_{j}\right\rangle -1\right)
  \label{mCDWdef}
\end{equation}
and
\begin{equation}
m_{SDW}=\frac{1}{L}\sum_{j}\left(-1\right)^{j}\left\langle \hat{s}_{j}^{z}\right\rangle.
\label{mSDWdef}
\end{equation}

As can be noticed in Figs.~\ref{EHM_phase_diagram}(a) and (d),
for repulsive interactions, the EHM features a CDW phase for $U<2V$ and a Mott insulator phase with SDW for $U>2V$. The color-codes correspond to the calculated values of $|m_{CDW}|$ [panel(a)] and $|m_{SDW}|$ [panel (d)]. These results are well known and in good agreement with early calculations~\cite{jeckelmann,EjimaEHM2007}. For small $U$ and $V$ around the $U=2V$ line, an additional instability occurs due to the competition between onsite and nearest-neighbor Coulomb interactions, giving rise to a BOW~\cite{nakamura2000tricritical}, which is not addressed in our work.


\subsection{Turning on either $U$ or $V$}\label{sec_results1}

Now we study the finite-time quench produced by turning on only one of the interactions: the onsite interaction, $U$, for a final state in the SDW phase, or the nearest-neighbor one, $V$, for a final state in the CDW phase. As explained previously, in both cases, we start from a metallic state [GS of Eq.~(\ref{eq_EHM}) with fixed $U_0=V_0=0$] and analyze the system evolution as the interactions increase linearly up to $U_f=6, V_f=0$ [\#1 vertical line in Fig.~\ref{EHM_phase_diagram}(a)] or $U_f=0, V_f=3$ [\#6 horizontal line in panel (d) of the cited figure].

As will become clear from our results discussed below, we can identify three regimes in the quench evolution: impulse, intermediate, and adiabatic ones. Since our main goal is the formation of the CDW or SDW ordering, we consider a quench is adiabatic (rigorously quasi-adiabatic) if (1) the fidelity calculated between evolved and equilibrium states [the precise definition of this quantity is given in Eq.~(\ref{fidelity})] at the end of the quench evolution is larger than $0.99$ and (2) the difference between quantities calculated from the evolved state and the equilibrium GS is smaller than the uncertainties of our numerical calculation. In this case, it is expected that the system hardly evolves after the quench.   

\subsubsection{Spin and charge profiles}

First, we look at the behavior of the local spin $\langle \hat{s}^z_i\rangle$ and charge $\langle \hat{n}_i\rangle$ mean values as a function of the chain site $i$, calculated from the states at $t=t_f$. These quantities are displayed for different ramping time scales on the middle and right panels of Fig.~\ref{EHM_phase_diagram}, respectively. For comparison, we also show the spin and charge profiles corresponding to the GS of $H(t=0)$ and $H(t=t_f)$. 

We observe that for a short quench time scale ($\tau_U=0.01$ or $\tau_V=0.01$) the system goes through a sudden quench, a regime that we identify as an impulse one. In this case the state is frozen at the starting one: the observables at $t=t_f$ are equal to the ones at $t=t_0$, as can be notice in Fig.~\ref{EHM_phase_diagram}(b),(c),(e), and (f).
In contrast, if the quench happens slowly ($\tau_V=12.0$ for the quench towards the CDW phase and $\tau_U=2.5$ for the quench to SDW), an adiabatic process takes place. In this case, we observe the formation of interchanging patterns in the charge [panel (c)] and spin [panel (e)] mean values, which closely follow the profile of the equilibrium GS with finite $V_f$ or finite $U_f$ and are characteristic of the CDW and SDW phases, respectively. Between these two extremes, we observe an intermediate regime, also displayed in the figure for comparison.

For completeness, we also show in Fig.~\ref{EHM_phase_diagram} the local magnetization for the quench towards the CDW phase [panel (b)] and the charge density for the quench towards the SDW phase [panel (f)]. For the latter, no charge fluctuations are expected in the SDW phase and $\left\langle \hat{n}_{i}\right\rangle =1$ since we consider a half-filled system. For the former, we find a residual, non-zero magnetization, as it also occurs in the regime where the interactions are turned off [first plot in panel (b)]. This is a result of the conservation of total magnetization and the fact that we study an odd length chain at half-filling - there is always a net magnetization equal to $1/2$ spreading along the chain.

\begin{figure}
\begin{centering}
\includegraphics[width=\linewidth]{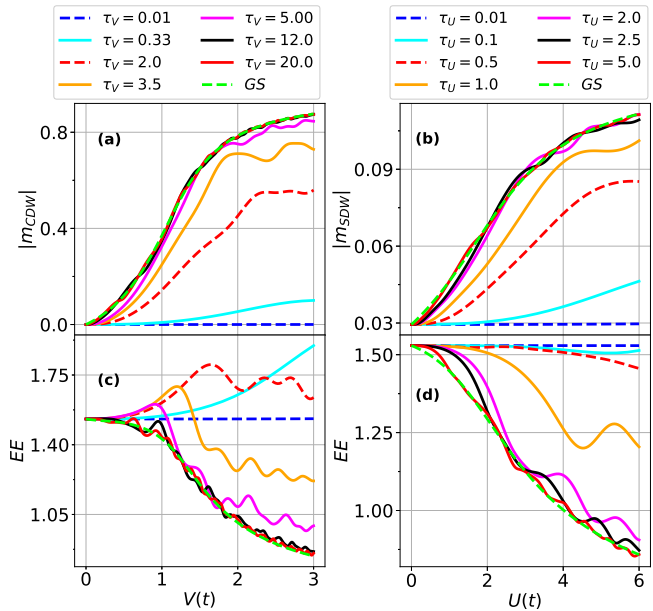}
\par\end{centering}
\caption{(a) CDW and (b) SDW order parameters and (c) and (d) the $EE$, all displayed along the quenches, that is, as a function of $V(t)$ for the quenches towards the CDW phase and as a function of $U(t)$ for the quenches aiming the SDW phase. \label{quench_mFS}}
\end{figure}

In Fig.~\ref{EHM_phase_diagram} we have plotted observables as a function of the chain site corresponding to the final state after the quench. Let us now look at observables defined for the whole system (that involve a sum over the sites, for example) and their evolution during the quench. 
In Figs.~\ref{quench_mFS}(a) and (b), we show the calculated order parameters as a function of the interactions, $V(t)$ or $U(t)$, for distinct ramping time scales, $\tau_V$ or $\tau_U$. In these figures, the green dashed lines represent the GS order parameters corresponding to the instantaneous equilibrium Hamiltonian at time $t$, that is, the Hamiltonian of Eq.~(\ref{eq_EHM}) with constant $U_{eq} = U(t)$ or $V_{eq} = V(t)$.

We observe that both $m_{CDW}$ and $m_{SDW}$ increase as the respective interaction increases. For our finite system, the former approaches the maximum value ($m_{CDW} \approx 1$) for $V_f=3$.
$m_{SDW}$, on the other hand, saturates at $\approx 0.11$ for $U_f= 6$. We recall that, in the limit of large $U$ and $V=0$, the half-filled system is equivalent to a Heisenberg chain~\cite{giamarchi2003quantum, altland2010condensed} with an exchange constant $\sim4J{{}^2}/U$. Even in this limit, we cannot observe a perfect N\'{e}el state because it is not an eigenstate of antiferromagnetic Heisenberg model~\cite{bauer2015temporal}. For this reason, in our problem, $m_{SDW}$ never reaches the maximum possible value (of $0.5$) allowed by the definition in Eq.~(\ref{mSDWdef}).

By comparing the different curves in Fig.~\ref{quench_mFS}(a), we conclude that $m_{CDW}$ calculated from the evolved state corresponding to $\tau_V=12.0$ approaches the equilibrium GS curve, revealing that $\tau_V \geq 12.0$ is necessary for our system to reach the adiabatic regime (we will come back to this point when discussing the fidelity in the next subsection).
In contrast, for $\tau_V=0.01$ the quench is sudden and the state does not evolve at all. The quenches with $0.01<\tau_V<12.0$ correspond to an intermediate regime; the order parameter increases but does not follow the GS value, indicating that the evolved state includes excited ones. 
This is clear in the case of $\tau_V=5.0$ (pink curve), for example, where we still observe some deviations between the results obtained from the evolved state and those of the equilibrium case.
Interestingly, the evolution of $m_{CDW}$ at the intermediate regime starts to exhibit some oscillatory behavior and is thus non-monotonic. 
The overall time evolution of $m_{SDW}$ [Fig.~\ref{quench_mFS}(b)] is similar to the one of $m_{CDW}$, however a smaller ramping time scale, $\tau_U=2.5$, is already enough to ensure the adiabatic regime in this case.
We can partially explain these different time scales looking at the interactions associated with the ordering of charges and spins along our quantum quenches. 
The rearrangement of charges towards the CDW ordering is dominated by the nearest-neighbor interaction $V$, while the rearrangement of spins towards SDW is governed approximately by $4J^{2}/U$ (see Fig.~\ref{EHM_phase_diagram}). 
As a consequence, the quenches towards the CDW ordering need to be longer than those towards SDW, in such a way that the excitations produced by the charge rearrangements are suppressed.


\begin{figure}
\begin{centering}
\includegraphics[width=\linewidth]{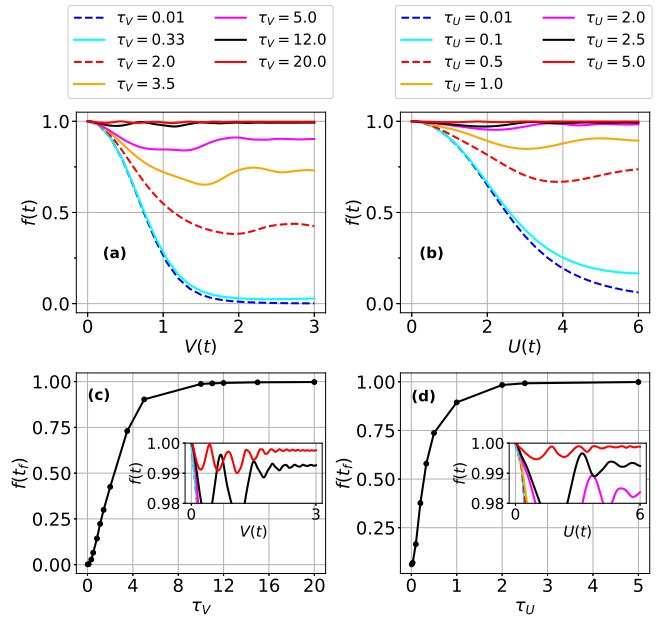}
\par\end{centering}
\caption{In (a) and (b) we present the fidelity $f(t)$ as a function of $V(t)$ and $U(t)$, respectively. In (c) and (d) we plot the fidelity at the end of the quench evolution $f(t_f)$ as a function of $\tau_V$ and $\tau_U$, respectively. The insets show a zoom of the data in panels (a) and (b).} \label{fig_fidelity}
\end{figure}

\subsubsection{Fidelity between evolved and equilibrium states}

To better characterize the observed regimes, we calculate the fidelity $f(t)$ between the evolved state at time $t$ and the GS corresponding to $U_{eq} = U(t)$ or $V_{eq} = V(t)$. $f(t)$ is defined as:
\begin{equation}
f(t)=\left|\left\langle \Psi(t)|\Psi_{GS}\right\rangle \right|^{2}, \label{fidelity}
\end{equation}
where $\left|\Psi\left(t\right)\right\rangle $ is the evolved state
and $\left|\Psi_{GS}\right\rangle $ the instantaneous GS.
We consider that when the fidelity at the end of the quench evolution remains larger than a threshold value
close to unity (e.g. $0.99$, with a tolerance of $0.01$), the process
can be viewed as adiabatic. This assures us that the system will follow the GS behavior in a free evolution after the quench since no additional excitations
were created by it. Otherwise, the evolved state includes excited ones and the post quench evolution is nontrivial.

In Fig.~\ref{fig_fidelity}(a) and (b), we display the calculated fidelity for the different regimes observed for quenches
towards the CDW and SDW phases, respectively.
In the impulse regime $(\tau_V=0.01$ and $\tau_U=0.01)$, we observe that the fidelity vanishes in the CDW phase ``earlier'' [around $V(t)=2$] than when the quench is towards the SDW phase, where $f(t)$ decays more slowly.
It happens because the equilibrium GS of the instantaneous Hamiltonian in the SDW portion of the phase diagram keeps a certain overlap with the initial state, at which $\left|\Psi\left(t\right)\right\rangle $ is frozen. 


For intermediate $\tau_U$ or $\tau_V$, the fidelity initially decreases and then stabilizes (with small oscillations) at a finite value, indicating that the evolved state is not orthogonal to the equilibrium GS. As the ramping time scale increases even further, we observe $f\thickapprox{\rm 1}$, characteristic of an adiabatic evolution. The behavior of the fidelity at the end of the quench evolution, $f(t_f)$, as a function of $\tau_V$ and $\tau_U$ is plotted in panels (c) and (d) of Fig.~\ref{fig_fidelity}. For the quench towards SDW ordering [see panels (b) and (d)], the process with $\tau_U=2.0$ is nearly adiabatic, while the one with $\tau_U=2.5$ is already in the adiabatic regime [$f(t_f)=0.992$]. For the quench towards the CDW ordering [panels (a) and (c)], on the other hand, we obtain $f(t_f)=0.903$ for $\tau_V=5.0$, which is still in the intermediate regime. The adiabatic evolution is reached only with a larger ramping time, $\tau_V=12.0$ [$f(t_f) = 0.996$], as already pointed out when we discussed the order parameter in the previous subsection.

We note that, for $\tau_V=12.0$, at the beginning of the quench, the fidelity is slightly smaller than our adiabatic criterium ($f>0.99$), with minor excitations created close to the critical point, however, the system equilibrates during the finite-time quench. Such a behavior is also obtained for $\tau_U=2.5$ in the quench towards the SDW phase, as can be seen in the insets of Fig.~\ref{fig_fidelity}, which depict a zoom in the region of $f\approx1$ of the data in panels (a) and (b). For larger $\tau$ (see $\tau_V=20$ and $\tau_U=5.0$), the fidelity remains above 0.99 during the whole quench. The differences between the evolution driven by the onsite or the nearest-neighbor interaction will be further explored below.


\subsubsection{Entanglement entropy}

To quantify how non-local correlations between parts of our system evolve along the applied quenches, we evaluate the bipartite entanglement entropy, defined as $EE=-\sum_{i}\lambda_{i}{\rm log}_{2}\lambda_{i}$, where the set $\left\{ \lambda_{i}\right\} $ is the so-called entanglement spectrum of eigenvalues of the reduce density matrix
$\hat{\rho}_{{\rm A}/{\rm B}}={\rm Tr}_{{\rm B}/{\rm A}}\left|\Psi\right\rangle
\left\langle \Psi\right|$. Here, we consider that the subsystem A contains the $(L-1)/2$ leftmost sites and B the $(L+1)/2$ rightmost one. We checked that the behavior of $EE$ observed by us does not qualitatively change if we take other subsystem sizes.


The evolution of the $EE$ throughout the quenches we consider is shown in Fig.~\ref{quench_mFS}(c) and (d). For comparison, we also display the $EE$ corresponding to the instantaneous equilibrium GS (green lines).
For the smallest values of the quench time ($\tau_U=0.01$ and $\tau_V=0.01$), that is, in the impulse regime, the entropy does not evolve from the respective initial values, those that correspond to $U_0 = 0$ or $V_0 = 0$. More interestingly, in quenches to the CDW phase, we observe an enhancement of the $EE$ for $\tau_V =0.33$, $\tau_V = 2.0$, $\tau_V = 3.5$, and $\tau_V = 5.0$ [cyan, red, orange, and magenta lines in Fig.~\ref{quench_mFS}(c)]. For $\tau_V=0.33$ the $EE$ increases monotonically during the quench. For $\tau_V = 2.0$, it increases up to $\approx 1.8$ and roughly saturates - in this case the CDW is not well-formed, as indicated by the small values of $|m_{CDW}|$ observed in panel (a). In the case of $\tau_V = 3.5$ and $\tau_V=5.0$, we observe a smaller enhancement of the $EE$, which is followed by a sudden suppression towards the GS behavior. These results indicate that during a quench in the intermediate time regime, before approximating the CDW ordering, the system access more entanglement excited states.  
On the other hand, we do not find any considerable enhancement of the $EE$ along the formation of the SDW phase, as can be noticed in Fig.~\ref{quench_mFS}(d). 
These observations suggest that the formation of the CDW ordering along a linear quench requires a larger rearrangement of our system than the formation of the SDW one. 
An increase of $EE$ close to a critical point was also observed in the dynamics of the Ising model with a time-dependent transverse field~\cite{Canovi2014}.

In the intermediate regime, we also observe oscillations in the $EE$ throughout the quench evolution - look at the results for $\tau_{V} = 3.5$ and $\tau_{U} = 1.0$ (orange lines). These oscillations decrease in amplitude as we approach the adiabatic regime. 
In this case, the $EE$ follows closely the results for the equilibrium GS state (green lines), as expected - both of them decrease linearly when the system goes from the metallic to the (ordered) insulating state.
Oscillations in the $EE$ as well as in the expectation values of operators that do not commute with the Hamiltonian were observed in spin chains after sweeping the Hamiltonian through a critical point~\cite{pollmann2010dynamics,Canovi2014}. These features can be viewed as consequences of the fact that the time evolved state includes excited states of the equilibrium Hamiltonian. 

\subsubsection{Deviations with respect to equilibrium ground state}

\begin{figure}
\begin{centering}
\includegraphics[scale=0.45]{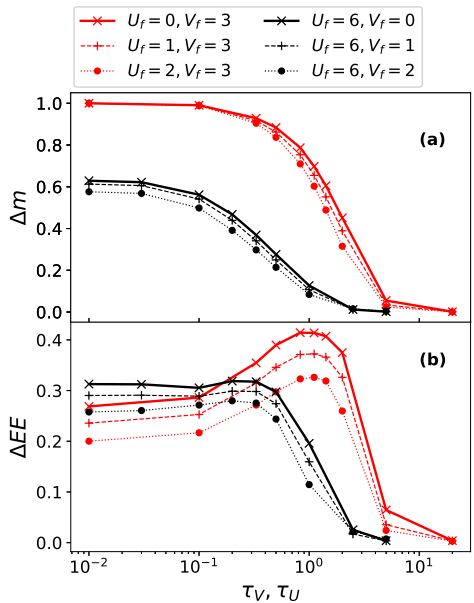}
\par\end{centering}
\caption{Deviations as defined in the text calculated for (a)~the CDW and SDW order parameters and
  (b)~the $EE$ as a function of the ramping times, $\tau_V$ for the quench towards the CDW phase (curves
  in red) and $\tau_U$ for the quench towards the SDW phase (curves in black). \label{quench_vel}
  }
\end{figure}

To summarize the main differences between the quenches towards the CDW and SDW phases, we define the deviation of a given quantity $\Theta$ as 
\begin{equation}
\Delta\Theta = \frac{\sum\left|\Theta_{t}-\Theta_{GS}\right|}{\sum\left|\Theta_{GS}\right|},
\end{equation}
where $\Theta_{t}$ and $\Theta_{GS}$ refer to the values of $\Theta$ evaluated with the evolved state and the equilibrium GS, respectively, and the sum is over the
instant of times between $t=0$ and $t=t_f$, that is, over the quench time evolution.
In Fig.~$\ref{quench_vel}$(a) we present our results for the deviations of $m_{CDW}$ and $m_{SDW}$ as a function of the respective quench ramping times, $\tau_V$ for the quenches towards the CDW phase and $\tau_U$ for quenches aiming the SDW phase. 
Here, we focus on the curves at which only one of the interactions, either $U$ or $V$, are turned on (solid lines);
the other results will be discussed in the next section.

On one hand, in both quench directions we observe that the order parameter deviations decrease monotonically as the ramping time increases. However, they vanish at a smaller $\tau$ when the quench is towards the SDW phase (black curves) than when it is towards the CDW phase (red curves), emphasizing that the adiabatic regime can be reached for smaller quench time scales in the former. On the other hand, the $EE$ deviations shown in Fig.~\ref{quench_vel}(b) features a non-monotonic behavior for the quench towards the CDW ordering, with enhanced deviations in the intermediate regime, compatible with the $EE$ behavior observed in Fig.~\ref{quench_mFS}(c). In the case of the quench towards the SDW ordering, practically no enhancement is observed in the EE, in agreement with Fig.~\ref{quench_mFS}(d). A comparison between the red (CDW) and black (SDW) curves emphasizes the fact that in the former excited states are accessed in the intermediate regime and thus the adiabatic case is reached with a larger ramping time.

\subsection{Turning on $U$ and $V$ simultaneously}\label{sec_results3}

We now investigate the effects of turning on both interactions, $U$ and $V$, simultaneously. At equilibrium, the Hubbard model [Eq.~(\ref{eq_EHM}) with $U(t)=U_{eq}\neq 0$ and $V(t)=V_{eq}=0$] can be exactly solved by the Bethe ansatz method~\cite{lieb2004absence}. Its extended version ($V_{eq}\neq0$), on the other hand, is non-integrable for general values of the model parameters~\cite{poilblanc1993poisson}. It was shown that integrable many-body quantum systems in one dimension could undergo relaxation to an equilibrium state described by the Generalized Gibbs ensemble~\cite{rigol2007relaxation} after a quench. The same does not apply to non-integrable models~\cite{d2016quantum,kormos2017real}. Such a difference has motivated us to address the effects of small $V$ values on the quench towards the SDW phase and small $U$ on the quench towards the CDW phase.  

In Fig.~\ref{quench_vel} we present the deviations of the different quantities when considering the quenches paths numbered 2, 3, 4, and 5 in panels (a) and (d) of Fig.~\ref{EHM_phase_diagram}. We compare them with the results for the vertical and horizontal quenches (paths 1 and 6) described in the previous section. We observe that the deviations of the order parameter and of the $EE$ decrease as $U_f$ increases for fixed $V_f=3$ (quenches towards the CDW phase) or $V_f$ increases for fixed $U_f = 6$ (quenches towards the SDW phase). More importantly, the three distinct regimes - sudden quench, intermediate, and adiabatic - are still clearly observed as a function of $\tau$ for the quenches with non-zero $U_f$ {\it and} $V_f$. In fact, we find very similar results to the ones discussed in the previous section, when only one interaction is turned on, indicating that the behavior of the evolved state during the finite-time quench does not depend on the integrability of our model.


\section{Conclusions}\label{sec_conclusion}

In this work we have performed time-dependent DMRG calculations to study the formation of CDW and SDW phases within the extended Hubbard model. To this aim, we have considered that our system is subjected to an interaction quench, that is, it is prepared in an initial non-interacting state and the interactions then change over a finite interval of time until their final values are reached. 

For the quenches we have analyzed,
three different quench regimes - impulse, intermediate, and adiabatic - are observed depending on the ramping quench time $\tau$. For small $\tau$, we have an impulse regime, in which the system remains frozen in the initial state.
In the intermediate regime, for the quench towards the CDW phase, we observe an increase of the entanglement entropy with respect to the initial value, not seen for the quench towards the SDW phase. This suggests that, during the time evolution towards the electronic CDW phase, more entanglement excited states of the equilibrium Hamiltonian are accessed, which does not happen during the formation of the SDW phase. 
As a consequence, we observe that the third regime, the adiabatic one, is reached with smaller ramping time scale if a SDW ordered state is formed as compared to the formation of a CDW state - the latter has to happen slowly to prevent entangled excited states from being accessed during the quench. 

Finally, our findings show that the breaking of the system integrability, produced by turning on the nearest-neighbor interaction $V$, does not induce significant changes in the non-equilibrium behavior during our quench.
We believe, however, that the non-integrability can affect the free evolution after the quench, especially for the states generated in the intermediate regime, since the system is excited at $t=t_f$. This effect can be investigated in future work.

\begin{acknowledgments}
The authors acknowledge financial support from the Brazilian agencies 
CAPES, CNPq (in particular CNPq INCT-IQ and grant 422300/2021-7), FAPDF, and FAPEMIG. 
\end{acknowledgments}

\appendix

\section{Dependence of the results on the system size} \label{sec_size}

\begin{figure}[!htb]
\begin{centering}
\includegraphics[scale=0.45]{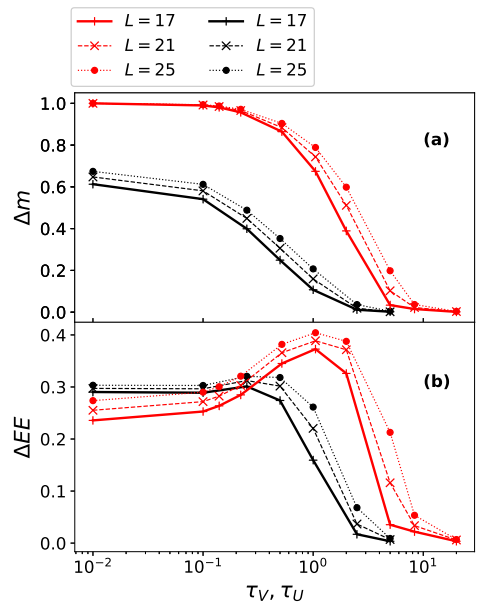}
\par\end{centering}
\caption{Deviations of (a) the order parameters and (b) the $EE$ as a function of the quench ramping time $\tau_V$ or $\tau_U$, depending on whether the quench is towards the CDW (red curves, for which $U_f = 1, V_f = 3$) or the SDW (black curves, for which $U_f = 6, V_f = 1$) phase. Results for different chain length $L$ are shown, including the one, $L = 17$, considered in the other plots. 
\label{DF_L_sizes}}
\end{figure}

To analyze how our results depend on the number of chain sites, we have selected the quench with $U_f = 1, V_f = 3$ [path \#2 in Fig.~\ref{EHM_phase_diagram}(a)] as representative of the formation of the CDW ordering and that with $U_f = 6, V_f = 1$ [path \#5 in Fig.~\ref{EHM_phase_diagram}(d)] for the case of the SDW ordering. 
In Fig.~$\text{\ref{DF_L_sizes}}$, we show the deviations of the order parameters and of the EE as a function of the quench time for different chain length $L$. As can be noticed in the figure, the increase in the number of sites does not qualitatively affect our findings. Moreover, before achieving the adiabatic behavior, we observe that chains with more sites present larger deviations of both quantities analyzed.
According to the adiabatic theorem, small energy gaps between the ground and first excited states lead to more excitations when the system crosses a critical point, which in our case occurs when the interactions are turned on (the system initial state coincides with the critical point). 
For systems with a finite number of sites, the gap decreases as the chain size increases~\cite{pollmann2010dynamics}, which leads to more excitations at the beginning of the quench evolution and thus to an increase in the deviations of quantities along the quench, as observed in our results. 

\bibliographystyle{apsrev4-2}
\bibliography{paper_hwc6}

\end{document}